\documentclass[11pt]{article}
\newtheorem{theorem}{Theorem}
\begin{document}

\title{Holonomy and projective symmetry in space-times\footnote{Submitted to Classical and Quantum Gravity, IOP publishing}}
\author{G S Hall\dag\ \& D P Lonie\ddag\ \\
Department of Mathematical Sciences, University of Aberdeen, \\
Meston Building, Aberdeen, AB24 3UE, Scotland, U.K. \\ \dag
e-mail: gsh@maths.abdn.ac.uk \ddag e-mail: DLonie@aol.com}

\maketitle

\begin{abstract}
It is shown using a space-time curvature classification and
decomposition that for certain holonomy types of a space-time,
proper projective vector fields cannot exist. Existence is
confirmed, by example, for the remaining holonomy types. In all
except the most general holonomy type, a local uniqueness theorem
for proper projective symmetry is established.
\end{abstract}

\section{Introduction}
Some techniques for studying projective symmetry in general
relativity theory were developed by the present authors in
\cite{ref1,ref2}. That study was heavily based on the algebraic
structure of the curvature tensor and included an account of the
relationship between the existence of (proper) projective
symmetry in a space-time and the latter's holonomy type. This
account unfortunately suffered from technical restrictions on the
`rank' of the curvature tensor and these restrictions were
removed in \cite{ref3}. A more natural and elegant account of this
relationship will be given here. To avoid undue repetition, the
results in \cite{ref1} will be used regularly here as will those
dealing with space-time holonomy theory in \cite{ref4,ref5}.

Throughout, $M$ will denote a (smooth) space-time manifold with
(smooth) Lorentz metric $g$ and $M$ will be assumed {\it
non-flat} in the sense that the associated curvature tensor (with
components $R_{abcd}$) does not vanish over any non-empty open
subset of $M$. The tangent space to $M$ at $m\in M$ is denoted by
$T_mM$ and a comma, a semi-colon and the symbol $\cal L$ are used
for a partial, a covariant and a Lie derivative respectively. A
global smooth (in fact $C^3$ is sufficient \cite{ref1}) vector
field $X$ on $M$ is called {\it projective} if its associated
local diffeomorphisms (flows) map geodesics into geodesics (not
necessarily preserving the affine parameter). This is equivalent
to the condition that, on decomposing the covariant derivative of
$X$ in any coordinate system in $M$ into its symmetric and
skew-symmetric parts as $X_{a;b}=\frac{1}{2}h_{ab}+F_{ab}$
($h_{ab}=h_{ba}$, $h={\cal L}_X g $, $F_{ab}=-F_{ba}$), $h$
satisfies \cite{ref6,ref7,ref1}
\begin{equation}
  h_{ab;c}=2g_{ab}\psi_c+g_{ac}\psi_b+g_{bc}\psi_a
\end{equation}
for some uniquely determined {\it closed} 1-form $\psi$ (the {\it
projective 1-form} of $X$) on $M$. If $h_{ab;c}=0$ on $M$
(equivalently, $\psi=0$ on $M$ \cite{ref1}), $X$ is affine (and
the associated local flows also preserve the affine parameter of
the geodesics of $M$) and, otherwise, $X$ is {\it proper
projective}. If $X$ is projective and satisfies $\psi_{a;b}=0$ on
$M$, $X$ is called {\it special projective}.

\section{Curvature decomposition}
In this section a review is given of a convenient classification
of the curvature tensor of $M$ \cite{ref8}. Let $m\in M$ and
consider the linear map $f$ from the 6-dimensional vector space
$B(m)$ of bivectors at $m$ into itself, given by $f:
F^{ab}\rightarrow R^{ab}_{\ \ cd}F^{cd}$. The curvature tensor at
$m$ can be put into exactly one of the following five disjoint
classes (and note that the class labelling differs from that in
\cite{ref1}).

\begin{description}
\item[Class $A$] This is the most general class and the curvature
will be said to be of this class if it is not one of the classes
$B$, $C$, $D$ or $O$ below.
\item[Class $B$] The curvature is said to be of class $B$ if the range of
$f$ is 2-dimensional and spanned by a timelike-spacelike pair of
simple bivectors with orthogonal blades.
\item [Class $C$] The
curvature is said to be of class $C$ if the range of $f$ is 2- or
3-dimensional and if there exists $k\in T_mM$, $k\neq 0$, such
that each bivector $F$ in this range satisfies $F_{ab}k^b=0$.
\item[Class $D$] The curvature is said to be of class $D$ if the range of
$f$ is 1-dimensional.
\item[Class $O$] The curvature is said to be of class $O$ if it vanishes at
$m$.
\end{description}

The dimension of the range of $f$ is just the {\it curvature
rank} when the curvature tensor $R_{abcd}$ is written in the
well-known way as a $6\times6$ symmetric matrix. The following
results are then straightforward to check (c.f. \cite{ref8}).
\begin{enumerate}
\item For the classes $A$ and $B$ there does {\bf not} exist
$k\in T_mM$, $k\neq 0$, such that $F_{ab}k^b=0$ for each $F$ in
the range of $f$. \item For class $A$, dim(range of $f$)$\geq2$
and if dim(range of $f$)$\geq4$ the class is necessarily $A$.
\item The vector $k$ in the definition of class $C$ is
unique up to a scaling. \item For class $D$ the identity
$R_{a[bcd]}=0$ shows that the range of $f$ is spanned by a {\it
simple} bivector.
\end{enumerate}

Let $V_m$ be a subspace of $B(m)$. If $F\in V_m$ is simple, its
blade is a 2-dimensional subspace of $T_mM$. If $F\in V_m$ is not
simple it may be written as $F=G+H$ where $G$ and $H$ are a
timelike-spacelike pair of simple bivectors with orthogonal
blades which are uniquely determined by $F$ \cite{ref9}. Now
define the subspace $U_m\subseteq T_mM$ associated with $V_m$ as
the span of the union of the blades of the members of $V_m$
(including both blades for a non-simple member of $V_m$). Thus if
$V_m$ is non-trivial, $\dim U_m\geq2$, if $V_m$ contains a
non-simple member $\dim U_m=4$ and if $\dim U_m<4$ each member of
$V_m$ is simple and there exists $k\in T_mM$, $k\neq0$, such that
$F_{ab}k^b=0$ for each $F$ in $V_m$. Also, if all members of $V_m$
are simple, $\dim V_m\leq3$. To see this, briefly, suppose $\dim
V_m\geq4$ with all members of $V_m$ simple. Let $F$ and $G$ be
independent members of $V_m$ so that $F$, $G$ and $F+\lambda G$
are simple for each $\lambda\in\mathrm{R}$. Now $E\in V_m$ is
simple if and only if ${E^a_{\ b}}^{*}\!E^b_{\ c}=0$, where *
denotes the duality operator. Applying this to the above three
bivectors yields ${F^a_{\ b}}^*\!G^{b}_{\ c}+{G^a_{\
b}}{^*\!F^{b}_{\ c}}=0$. Now, since $F$ is simple, there exists
$k\in T_mM$, $k\neq0$, such that $F_{ab}k^b=0$ and a contraction
of the previous equation with $k_a$ gives $(k_aG^a_{\
b})^*\!F^b_{\ c}=0$. Thus the vector $G^a_bk^b$, which is in the
blade of $G$, is either zero or orthogonal to the blade of the
(necessarily simple) bivector $^*\!F$ and hence in the blade of
$F$. In either case, the blades of $F$ and $G$ (and of $^*\!F$ and
$^*\!G$) intersect non-trivially and so there exists $k'\in
T_mM$, $k'\neq0$, such that $F_{ab}k'^b=G_{ab}k'^b=0$. Thus,
adopting a standard notation in which a simple bivector whose
blade is spanned by $r,s\in T_mM$ is written $r\wedge s$, one has
$p,x,y\in T_mM$ such that $F=p\wedge x$ and $G=p\wedge y$. Now
let $H\in V_m$ with $F$, $G$ and $H$ independent and apply the
above argument to the pairs $(F,H)$ and $(G,H)$ to see that
either $H=p\wedge z$ ($z\in T_mM$) or that the above vectors $x$
and $y$ may be chosen so that $H=x\wedge y$. Finally, introduce
$K\in V_m$ with $F$, $G$, $H$ and $K$ independent and apply a
similar argument to $F$, $G$ and $K$ to see that one contradicts
either the independence of $F$, $G$, $H$ and $K$ or the fact that
all members of $V_m$ are simple. This completes the proof. It
follows from the first part of this proof that if all members of
$V_m$ are simple and $\dim V_m=2$, there exists $k\in T_mM$,
$k\neq0$, such that $F_{ab}k^b=0$ for each $F$ in $V_m$, and
$\dim U_m=3$.

Now identify the range of $f$ with $V_m$ above. Then, if the
curvature class at $m$ is $D$, $\dim V_m=1$ and $\dim U_m=2$
whereas if it is $C$, $\dim V_m=2$ or $3$ and $\dim U_m=3$ and
for class $B$ one has $\dim V_m=2$ and $\dim U_m=4$. If $\dim
U_m\leq3$ the curvature class at $m$ is $O$, $C$ or $D$. It can
also now be checked by using `rank type' arguments that there
exists an open neighbourhood $W$ of $m$ such that $\dim
V_{m'}\geq\dim V_m$ and $\dim U_{m'}\geq\dim U_m$ for each $m'\in
W$.

Denote by the same symbols $A$, $B$, $C$, $D$ and $O$ the subsets
of $M$ consisting of precisely those points where the curvature
is of that class. Then $M=A\cup B\cup C\cup D\cup O$ and, since
$M$ is non-flat, $O$ has empty interior in the usual manifold
topology on $M$. The following theorem can now be proved (and an
independent proof can be found in \cite{ref10}).
\begin{theorem}
$M$ may be disjointly decomposed as \[M=A\cup \mathrm{int}B\cup
\mathrm{int}C\cup \mathrm{int}D\cup Z\] where $\mathrm{int}$
denotes the topological interior operator in $M$, $A$ is open,
$Z$ is a closed subset of $M$ defined by the disjointness of the
decomposition and where $\mathrm{int}Z=\emptyset$ (and so
$M\backslash Z$ is an open dense subset of $M$).
\end{theorem}

{\it Proof}. This will be briefly sketched making free use of the
above comments together with the usual rank theorems. Let $m\in
A$ so that $\dim V_m\geq2$ and $\dim U_m=4$. Then there exists an
open neighbourhood $N_1$ of $m$ such that $\dim U_{m'}=4$ for
each $m'\in N_1$ and so the curvature class at each point of
$N_1$ is $A$ or $B$. If $\dim V_m\geq3$, one may choose $N_1$
such that $\dim V_{m'}\geq3$ for each $m'\in N$ and so
$N_1\subseteq A$. If $\dim V_m=2$, and since (from remarks above
because $m\in A$) $V_m$ must contain a non-simple member, it may
be arranged that $V_m$ is spanned by non-simple bivectors $F$ and
$G$. But the blade-pairs of $F$ and $G$ cannot coincide at $m$
(since then $m\in B$) and so they will not coincide over some open
neighbourhood $N_2$ of $m$. Then with the original $N_1\subseteq
A\cup B$, $N_1\cap N_2\subseteq A$ and $A$ is open. Now let $m\in
B$ so that $V_m$ contains a non-simple member. Then $V_{m'}$
contains a non-simple member for each $m'$ in some neighbourhood
$N_3$ of $m$ and so $N_3\subseteq A\cup B$ and so $A\cup B$ is
open. A simple consideration of rank on $V_m$ shows that $A\cup
B\cup C$ and $A\cup B\cup C\cup D$ are also open. Finally, to show
$\mathrm{int}Z=\emptyset$, let $W\subseteq Z$ be open. Then, by
disjointness, $W\cap A=\emptyset$. Then, since, $A\cup B$ is
open, $W\cap B=W\cap(A\cup B)$ is open. But $Z$ is disjoint from
$\mathrm{int}B$ and so $W\cap B=\emptyset$. Similarly, $W\cap
C=W\cap D=W\cap O=\emptyset$ and so $W=\emptyset$ and
$\mathrm{int}Z=\emptyset$. $\bullet$

\section{Holonomy structure}
A holonomy classification scheme for a space-time $M$ was given
in \cite{ref4,ref5}. There it was described how, when $M$ is
simply connected, the holonomy group of $M$ (which is a connected
Lie group) is determined by its Lie algebra, this latter being a
subalgebra of the Lie algebra of the Lorentz group. The holonomy
type was then labelled $R_1$,...,$R_{15}$ following a similar
labelling of the subalgebras of the Lorentz algebra given in
\cite{ref11} with $R_1$ being the trivial case when $M$ is flat
and $R_{15}$ the most general. It is noted here for future
reference that the dimensions of the non-trivial holonomy
algebras are $R_2$, $R_3$, $R_4$ (dimension 1), $R_6$, $R_7$,
$R_8$ (dimension 2), $R_9$, $R_{10}$, $R_{11}$, $R_{12}$,
$R_{13}$ (dimension 3), $R_{14}$ (dimension 4) and $R_{15}$
(dimension 6). The type $R_5$ cannot occur as the holonomy group
of a space-time (see, e.g. \cite{ref4}). Another useful result
(from infinitesimal holonomy theory) is that, in the notation of
the last section, the range of $f$ at each $m\in M$ is in an
obvious sense, using the bivector representation of the Lorentz
algebra, contained in the holonomy algebra \cite{ref12}. Thus if
the holonomy type of $M$ is $R_1$, the curvature class at any
point of $M$ is $O$ whereas for types $R_2$, $R_3$ and $R_4$ it
is $O$ or $D$, for $R_6$ and $R_8$ it is $O$, $D$ or $C$, for
$R_7$ it is $O$, $D$ or $B$, for $R_9$ and $R_{12}$ it is $O$,
$D$, $C$ or $A$, for $R_{10}$, $R_{11}$ and $R_{13}$ it is $O$,
$D$ or $C$ and for $R_{14}$ and $R_{15}$ it could be any
curvature class. A table which gives, for each holonomy type, the
covariantly constant and recurrent vector fields admitted by $M$
can be found in \cite{ref5}.

In \cite{ref1} some results regarding the incompatibility of the
existence of {\it proper} projective vector fields on $M$ and
certain holonomy types for $M$, together with the assumed
constancy of the dimension of the range of the linear map $f$,
were given. In the next section these assumptions will be removed
with the help of theorem 1. The role played by the map $f$ is
also crucial and, in particular, the interplay between its range
and kernel where the latter consists of those bivectors $G$
satisfying $R_{abcd}G^{cd}=0$.

\section{The main theorems}
Suppose $X$ is a projective vector field on $M$ as in section 1.
Suppose also that some bivector $G$ lies in the kernel of the map
$f$. Then it was shown in \cite{ref1} that the following
relations hold;
\begin{equation}
h_{ae}R^e_{\ bcd}+h_{be}R^e_{\
acd}=g_{ac}\psi_{b;d}-g_{ad}\psi_{b;c}+g_{bc}\psi_{a;d}-g_{bd}\psi_{a;c}
\end{equation}
\begin{equation}
\psi_{a;c}G^c_{\ b}+\psi_{b;c}G^c_{\ a}=0
\end{equation}
These two equations dovetail in the following way. Condition (3)
at $m\in M$ is equivalent to the statement that, if $G$ is
simple, all non-zero vectors in the blade of $G$ are eigenvectors
of $\psi_{a;b}$ with equal eigenvalue, and if $G$ is non-simple,
the same condition holds for each of its (uniquely determined
orthogonal pair of) blades, but where the eigenvalue may depend on
the blade \cite{ref8}. Thus, if the kernel of $f$ is
`sufficiently' large, each member of $T_mM$ will be an
eigenvector of $\psi_{a;b}$ with the same eigenvalue and so, at
$m$, $\psi_{a;b}=\alpha g_{ab}$ ($\alpha\in\mathbf{R}$). But this
last condition on $\psi_{a;b}$ is equivalent to the right hand
side of (2) vanishing and hence to the vanishing of the left hand
side of (2). Finally, the vanishing of the left hand side of (2)
is the condition that (loosely stated) $h$ may be an `alternative
metric' \cite{ref8,ref13} and its solution for $h$ given the range
of the map $f$ is known \cite{ref8,ref14}. In particular, if the
range of $f$ satisfies the class $A$ condition at $m$, $h$ is
proportional to $g$ at $m$. Using techniques such as these it was
shown in \cite{ref1} that if $X$ is projective on $M$, the
algebraic structure of the curvature tensor at points of the
subsets $B$, $C$ and $D$ of $M$ resulted in the projective 1-form
$\psi$ vanishing on $\mathrm{int}B$ and $\mathrm{int}C$ and
satisfying $\psi_{a;b}=0$ on $\mathrm{int}D$. These results will
be useful in proving the next two theorems.
\begin{theorem}
Let $M$ be a non-flat simply connected space-time of holonomy
type $R_2$, $R_3$, $R_4$, $R_6$, $R_7$, $R_8$ or $R_{12}$. Then
$M$ does not admit a proper projective vector field.
\end{theorem}
{\it Proof}. Let $X$ be a projective vector field on $M$ and let
$M$ be decomposed as in theorem 1. If
$\mathrm{int}B\neq\emptyset$ let $m\in\mathrm{int}B$ and let
$W\subseteq\mathrm{int}B$ be an open, connected, simply connected
neighbourhood of $m$. Regarding $W$ as a space-time with metric
given by the restriction of $g$ to $W$, it follows (as remarked
above) from theorem 5 in \cite{ref1} that $\psi=0$ on $W$ and
hence on $\mathrm{int}B$. Similarly, theorem 4 in \cite{ref1}
shows that, if $\mathrm{int}C\neq\emptyset$, $\psi=0$ on
$\mathrm{int}C$. Now let $m\in A$. The assumption in the
statement of the theorem regarding the possible holonomy types of
$M$ and the remarks at the end of the first paragraph of section 3
show that $\dim(\mathrm{range \ of} f)\leq3$ and hence that the
kernel of $f$ has dimension at least three. The same assumption
and the fact that the range of $f$ is `contained' (in an obvious
sense) in the holonomy algebra enables the kernel of $f$ to be
calculated at $m$. This calculation easily reveals that this
kernel is sufficient to establish, using (3) as described above,
that $\psi_{a;b}=\gamma g_{ab}$ for some function
$\gamma:A\rightarrow\mathbf{R}$. Thus, as explained at the
beginning of this section, the right and hence left hand side of
(2) is zero and the vanishing of the left hand side shows that
$h$ is proportional to $g$ at $m$ and hence that, on $A$,
$h_{ab}=\beta g_{ab}$ for some function
$\beta:A\rightarrow\mathbf{R}$. A substitution of this into (1)
and comparing a contraction of (1) with $g^{ab}$ to a contraction
with $g^{ac}$ then shows that $\psi=0$ on $A$. Now let
$m\in\mathrm{int}D$ and with $W\subseteq\mathrm{int}D$ as above.
Then it follows (theorems 7 and 9 in \cite{ref1}) that, if $X$ is
proper, the holonomy type of $W$ is $R_{10}$, $R_{11}$ or
$R_{13}$ and a contradiction follows since these can not be
contained in the assumed holonomy groups for $M$. Hence $\psi=0$
on $\mathrm{int}D$. Thus $\psi=0$ on the open dense subset
$M\backslash Z$ of $M$ and hence on $M$ and the result follows.
$\bullet$

\begin{theorem}
Let $M$ be a non-flat simply connected space-time which admits a
proper projective vector field $X$. Then $M$ has holonomy type
$R_{10}$, $R_{11}$ or $R_{13}$ if and only if $X$ is special
projective.
\end{theorem}
{\it Proof}. If $M$ has any of these holonomy types and since the
range of $f$ at any $m\in M$ is contained in the holonomy algebra,
it follows from result (i) following the classification in
section 2 that the subsets $A$ and $B$ in the decomposition of
$M$ in theorem 1 are empty. Thus $M=C\cup\mathrm{int}D\cup Z$
(since, from rank considerations, $C$ is now open). If $X$ is a
proper projective vector field on $M$, it follows from theorem 4
in \cite{ref1} that $\psi=0$ on $C$ and from theorem 3 in
\cite{ref1} that $\psi_{a;b}=0$ on $\mathrm{int}D$. Thus
$\psi_{a;b}=0$ on the open dense subset $M\backslash Z$ of $M$
and hence on $M$ and so $X$ is special projective. [Also, since
$X$ is proper and $\psi$ covariantly constant, $\psi$ cannot be
zero at any $m\in M$ and so $C=\emptyset$ and $M=D\cup Z$ with
$D$ open in $M$.] Conversely, if $M$ admits a proper special
projective vector field $X$, then from section 7 in \cite{ref1}
one has a decomposition $M=D\cup Z$ and theorem 6 in \cite{ref1}
shows that $M$ admits a single independent covariantly constant
vector field (represented by $\psi^a$). From the possible
holonomy types available in this case, and recalling the
exclusions of theorem 2, it follows that the holonomy type of $M$
is $R_{10}$, $R_{11}$ or $R_{13}$. $\bullet$

If the conditions and conclusions of theorem 3 hold, the local
form of the metric in $M\backslash Z$ is known \cite{ref1}. Thus
theorems 2 and 3 essentially complete the study of (proper)
projective symmetry in space-times except in the cases of
holonomy type $R_9$, $R_{14}$ and $R_{15}$.

\section{Space-times of holonomy type $R_9$ and $R_{14}$}
A consideration of space-times of non-zero constant curvature
\cite{ref15} and certain F.R.W. models \cite{ref16} show that
proper projective vector fields exist in holonomy type $R_{15}$
space-times. In this section a brief discussion of space-times of
holonomy type $R_9$ and $R_{14}$ will be presented and which will
reveal a local uniqueness result for and the existence of proper
projective symmetry.

Let $M$ be a simply-connected space-time of holonomy type $R_9$
or $R_{14}$ and let $X$ be a proper projective vector field on
$M$ with associated 1-form $\psi$. Then $M$ admits a global,
null, nowhere-zero recurrent vector field $l$ (i.e. in each local
coordinate domain of $M$, $l_{a;b}=l_ap_b$ for some global
covector field $p$). On differentiating this relation and using
the Ricci identity one finds $l_dR^d_{\ abc}\!=\!l_aG_{bc}$, where
$G_{ab}=2p_{[a;b]}$, and then the identity $R^a_{\ [bcd]}=0$
leads to $G_{[ab}l_{c]}$ and hence to $G_{ab}l^b=\gamma l_a$ for
some function $\gamma:M\rightarrow\mathbf{R}$. Now let $p\in M$
and $U$ an open neighbourhood of $p$ on which $l$ has been
extended to a smooth null tetrad $l,n,x,y$ (whose only
non-vanishing inner products are $l^an_a=x^ax_a=y^ay_a=1$). The
fact that the range of $f$ must be contained in the holonomy
algebra means, from the holonomy algebras in the $R_9$ and
$R_{14}$ cases \cite{ref1}, that, in $U$, this range must be a
subspace of the span of the bivectors $l\wedge n$, $l\wedge x$,
$l\wedge y$ and $x\wedge y$ in the $R_{14}$ case and of the span
of the first three of these in the $R_9$ case. Thus if $H$ is
either of the bivectors $l\wedge x$ and $l\wedge y$,
$R_{abcd}H^{cd}=0$ and a contraction of (2) with $H^{cd}=0$ gives
\begin{equation}
H_a^{\ d}\psi_{b;d}+H_b^{\ d}\psi_{a;d}=0
\end{equation}
It follows from \cite{ref8} that $l$, $x$ and $y$ are
eigenvectors of $\psi_{a;b}$ at each $p\in U$ with equal
eigenvalue and hence that, on $U$,
\begin{equation}
\psi_{a;b}=\alpha g_{ab}+\beta l_al_b
\end{equation}
for smooth functions $\alpha$ and $\beta$ on $U$. Next one
differentiates (5) and uses the Ricci identity
$2\psi_{a;[bc]}=\psi_dR^d_{\ abc}$ to get
\begin{equation}
\psi_dR^d_{\
abc}=2g_{a[b}\alpha_{,c]}+2l_a\left(l_{[b}\beta_{,c]}+2l_{[b}p_{c]}\right)
\end{equation}
A contraction of (6) with $l^b$ then gives, on $U$,
\begin{equation}
G_{da}\psi^dl_c=l_a\alpha_{,c}-(\alpha_{,b}l^b)g_{ac}-(\beta_{,b}l^b+2\beta
p_bl^b)l_al_c
\end{equation}
A contraction of (7) with $x^ax^c$ then gives $\alpha_{,b}l^b=0$
and a back substitution into (7) followed by a contraction with
$n^a$ reveals that $\alpha_{,c}$ is proportional to $l_c$ on $U$.
A contraction of (6) with $l^a$ then gives $(\psi_dl^d)G_{bc}=0$
on $U$.

Now consider the open subset $A\subseteq M$ in the decomposition
of theorem 1 and in which the results of the previous paragraph
hold. In addition, one now has (result (i) in section 2) that $G$
is nowhere zero on $A$ and so $\psi_al^a=0$ on $A$. Thus
$(\psi_al^a)_{;b}=0$ and use of the recurrence condition on $l$
together with (5) then gives $\alpha=0$ and so $\psi_{a;b}=\beta
l_al_b$ on $A$. [From this it can also be shown that $\psi$ is
proportional to $l$ on $A$ but this fact will not be required.]
Now let $X$ and $Y$ be projective vector fields on $M$ with
associated 1-forms $\psi$ and $\phi$ and with (1) also holding
with $h$ and $\psi$ for $X$ replaced with $H$ and $\phi$ for $Y$.
Then, on $A$, $\psi_{a;b}=\beta l_al_b$ and $\phi_{a;b}=\delta
l_al_b$ for smooth functions $\beta$ and $\delta$ on $A$ and on
writing down (2) for $X$ and $Y$ one easily sees that the
associated right hand sides are proportional and hence
\begin{equation}
\tilde{h}_{ea}R^e_{\ bcd}+\tilde{h}_{eb}R^e_{\ acd}=0
\hspace{1cm}(\tilde{h}=\delta h-\beta H)
\end{equation}
Now it follows from \cite{ref8,ref14} that, on $A$, the only
solutions to (8) are of the form $\tilde{h}_{ab}=\nu g_{ab}$ for
some smooth function $\nu$ on $A$ and hence that $\delta
h_{ab}-\beta H_{ab}=\nu g_{ab}$. Now let $V\subseteq A$ be the
subset of A on which $\beta$ \underline{and} $\delta$ vanish. If
$\mathrm{int}V\neq\emptyset$ (and note that since $A$ is open in
$M$, it makes no difference whether the interior is taken in $M$
or in the subspace topology in $A$) then $\psi_{a;b}=0$ and
$\psi_dR^d_{\ abc}=0$ and so $\psi_{a}\equiv0$ on $\mathrm{int}V$
(result (i) section 2 again). Similar remarks apply to $\phi_a$
and so $X$ and $Y$ are affine on $\mathrm{int}V$. Now consider
the open subset $W\equiv A\backslash V$ of $M$. If $p\in W$ then
at least one of $\beta$ and $\delta$ is non-zero at $p$ and hence
in some open \underline{connected} neighbourhood $U(\subseteq W)$
of $p$. Suppose it is $\delta$ (a similar argument applies if it
is $\beta$) so that, on $U$, one may write $h=\rho H+\sigma g$ for
smooth functions $\rho$ and $\sigma$ on $U$. Then (1) gives on $U$
\begin{equation}
2g_{ab}\chi_c+g_{ac}\chi_b+g_{bc}\chi_a=H_{ab}\rho_{,c}+g_{ab}\sigma_{,c}
\hspace{1cm}(\chi=\psi-\rho\phi)
\end{equation}
Then, at any $q\in U$, contract (9) with a non-zero vector $r^c$
chosen to be orthogonal to $\chi_c$, $\rho_{,c}$ and
$\sigma_{,c}$, giving $2r_{(a}\chi_{b)}=0$, and hence $\chi_c=0$
at $q\in U$. It follows that $\chi_a$ vanishes on $U$. So let
$E\subseteq U$ be the open subset of $U$ (and of $M$) on which
$\rho_{,c}$ does not vanish. Then (9) shows that, on $E$,
$H_{ab}=\mu g_{ab}$ for some smooth function $\mu$ on $E$ and so,
from (1) applied to $H$ and $\phi$, one easily finds that if
$E\neq\emptyset$, $\phi_a$ and hence $\phi_{a;b}$ vanish on $E$.
This contradicts the fact that $\delta$ never vanishes on $U$ and
so $E=\emptyset$. So $\rho_{,a}$ vanishes on $U$
($\Rightarrow\rho$ is constant on $U$ since $U$ is connected) and
then $Z\equiv X-\rho Y$ is affine on $U$ since it is projective
on $U$ with vanishing projective 1-form $\chi$.

To summarise the previous paragraph, given any point $p$ in the
open dense set $\mathrm{int}V\cup W$ of $A$ there exists an open
neighbourhood of $p$ in which some linear combination of $X$ and
$Y$ is affine. [In fact, since one is working on a
\underline{subset of $A$}, it is known that proper affine vector
fields cannot be admitted on (any open subset of) this subset and
hence this linear combination of $X$ and $Y$ must be
\underline{homothetic}.] Also the restrictions of $X$ and $Y$ to
$\mathrm{int}B$ and $\mathrm{int}C$ in the decomposition of $M$
in theorem 1 are necessarily affine (as remarked in the proofs of
theorems 2 and 3) and for each $p\in \mathrm{int}D$ there is a
connected open neighbourhood $U$ of $p$ on which some linear
combination of $X$ and $Y$ is affine \cite{ref1}. Thus, in the
notation of the previous paragraph one may disjointly decompose
$A$ as $A=W\cup\mathrm{int}V\cup Z'$ where $Z'$ is defined by the
disjointness of the decomposition and is easily shown to have
empty interior in the subspace topology on $A$ from $M$ and hence
in the topology of $M$, $\mathrm{int}Z'=\emptyset$. Then one has
the disjoint decomposition
$M=W\cup\mathrm{int}V\cup\mathrm{int}B\cup\mathrm{int}C\cup\mathrm{int}D\cup
Z\cup Z'$ and some elementary topology reveals that $Z\cup Z'$ is
closed with empty interior and so $\tilde{M}\equiv
W\cup\mathrm{int}V\cup\mathrm{int}B\cup\mathrm{int}C\cup\mathrm{int}D$
is open and dense in $M$. One then has the following local
uniqueness result.
\begin{theorem}
Let $M$ be a non-flat simply connected space-time of holonomy
type $R_9$ or $R_{14}$. Then if $X$ and $Y$ are projective vector
fields on $M$ there exists an open dense subset $\tilde{M}$ of
$M$ such that each $p\in \tilde{M}$ admits an open neighbourhood
$U$ such that some linear combination of the restrictions of $X$
and $Y$ to $U$ is affine on $U$.
\end {theorem}
Consider the metric given in a global chart domain $M$ with
coordinates $(x^1,x^2,x^3,x^4)=(u,v,x,y)$ with $0<u,v<\infty$,
$-\infty<x,y<\infty$, by
\begin{equation}
ds^2=2dudv+\sqrt{uv^{-3}}dv^2+v^2e^{g(x,y)}(dx^2+dy^2)
\end{equation}
for some smooth function $g$ on $\mathbf{R}^2$. The only
non-vanishing components (up to algebraic symmetries) of the
curvature tensor are
\begin{equation}
R_{1212}=\frac{1}{8w^3}\hspace{6mm}R_{2323}=R_{2424}=\frac{-e^{g(x,y)}}{4w}\hspace{6mm}
R_{3434}=-\frac{v^2}{2}e^{g(x,y)}\nabla^2g
\end{equation}
where $w=\sqrt{uv}$ and
$\nabla^2g\equiv\frac{\partial^2g}{\partial
x^2}+\frac{\partial^2g}{\partial y^2}$. The vector field $l$ on
$M$ defined by $l^a=(1,0,0,0)$ is null and recurrent on $M$ and,
since it is easily checked from (11) that $R_{abcd}l^d$ never
vanishes on $M$, $l$ is not covariantly constant over any
non-empty subset of $M$. The existence of such a vector field $l$
together with the facts that the rank of the bivector map $f$ at
$p\in M$ is either 4 or 3 (the latter occurring if and only if
$(\nabla^2g)_p=0$) and that at each $p\in M$ the range of $f$ can
be spanned by simple bivectors shows that the holonomy type of the
metric (10) is either $R_9$ or $R_{14}$ \cite{ref5}. It is
clearly type $R_{14}$ if $f$ has rank 4 at some point of $M$ and,
less obviously, type $R_{9}$ if $\nabla^2g$ vanishes identically
on $M$. To see this last result one notes that if
$\nabla^2g\equiv0$ on $M$, and with $p\in M$ fixed, the range of
$f$ at any $p'\in M$ is spanned by the bivectors
$\frac{\partial}{\partial u}\wedge\frac{\partial}{\partial x}$,
$\frac{\partial}{\partial u}\wedge\frac{\partial}{\partial y}$
and $\frac{\partial}{\partial u}\wedge\frac{\partial}{\partial
v}$ (and $l=\frac{\partial}{\partial u}$) at $p'$. On parallel
transport of these bivectors from $p'$ along some curve $c$ to
$p$ the bivectors accumulated at $p$, for all points $p'$ and all
associated curves $c$, span the holonomy algebra of $M$ by the
Ambrose-Singer theorem \cite{ref17} (see also \cite{ref5}). Since
$l$ is recurrent (and thus its \underline{direction} is unchanged
under parallel transport) and since the null and simple
properties of bivectors are also, respectively, unchanged under
parallel transport, it is easily checked that the set of bivectors
accumulated at $p$ is the span of the above set of three
bivectors at $p$ and so the holonomy type is $R_9$. Finally, the
vector field $X$ given by $X^a=(uv,v^2,0,0)$ can now be checked
to be a \underline{proper} \underline{projective} vector field on
$M$ with associated 1-form $\psi$ given by $\psi=(0,1,0,0)$
independently of the rank of $f$ and thus confirms the existence
of proper projective vector fields in both $R_9$ and $R_{14}$
holonomy types.


\end{document}